\documentclass[10pt, conference, compsocconf]{IEEEtran}

\usepackage{color}
\usepackage{multirow}
\usepackage{tabularx}
\usepackage{array}
\usepackage{amssymb}

\ifCLASSINFOpdf
\else
\fi

\usepackage{graphicx}
\usepackage{amsmath}
\usepackage{authblk}
\usepackage{textcomp}
\graphicspath{{Figure/}}

\begin{document}
%
\title{Large Scale GPU Accelerated PPMLR-MHD Simulations for Space Weather 
Forecast}


\author[1]{Xiangyu Guo}
\author[2]{Binbin Tang\thanks{B.B@university.edu}}
\author[3]{Jian Tao\thanks{D.D@university.edu}}
\author[2]{Zhaohui Huang\thanks{C.C@university.edu}}
\author[1]{Zhihui Du*\thanks{D.D@university.edu}}
\affil[1]{Tsinghua National Laboratory for Information Science and Technology}
\affil[1]{Department of Computer Science and Technology,Tsinghua University, 100084, Beijing, China}
\affil[1]{*Corresponding Author's E-mail: duzh@tsinghua.edu.cn}
\affil[2]{State Key Laboratory of Space Weather, National Space Science Center, CAS}
\affil[3]{Center for Computation \& Technology, Louisiana State University, Baton Rouge, LA, USA}

\maketitle

\begin{abstract}
PPMLR-MHD is a new magnetohydrodynamics (MHD) model used to simulate the 
interactions of the solar wind with the magnetosphere, which has been proved to 
be the key element of the space weather cause-and-effect chain process from the 
Sun to Earth.  Compared to existing MHD methods, PPMLR-MHD achieves the 
advantage of high order spatial accuracy and low numerical dissipation. 
However, the accuracy comes at a cost. On one hand, this method requires more intensive computation. On the other hand, more boundary data is subject to be transferred during the process of simulation.
In this work, we present a parallel hybrid solution of the PPMLR-MHD model
implemented using the computing capabilities of both CPUs and GPUs. We 
demonstrate that our optimized implementation alleviates the data transfer overhead by using GPU Direct technology and can scale up to 151 processes and 
achieve significant performance gains by distributing the workload 
among the CPUs and GPUs on Titan at Oak Ridge National Laboratory. 
The performance results show that our implementation is fast 
enough to carry out highly accurate MHD simulations in real time.

\end{abstract}


%
\IEEEpeerreviewmaketitle

\section{Introduction}

The magnetosphere, which is the outermost of the geospace, is formed when the 
solar wind interacts with the Earth's internal magnetic field 
\cite{hu2007ionospheric}. Understanding the formation and development of the 
magnetosphere is crucial because it is the key element of the space weather 
cause-and-effect chain process from the Sun to the Earth. Similar to the weather 
on Earth, space weather is about the time varying conditions taking place 
in the space from the solar atmosphere to the geospace. It is driven by the 
solar wind which carries solar energy and comes through interplanetary space 
from the near surface of the Sun and the Sun's atmosphere. Due to the effect of 
the Earth's magnetic field, the magnetosphere forms a shape similar to a bullet, 
where the sunny side is roughly like a hemisphere with a radius of 15$R_{E}$(Earth 
radii), while the nightside forms a cylinder shape with a radius of 20-25$R_{E}$. 
The tail region stretches well beyond 200$R_{E}$ while its exact length is not well 
known. It is known that the geospace, including the magnetosphere and 
ionosphere, has the nature of nonlinearity, multiple-component, and 
time-dependence, which together pose a big challenge to investigate it using only 
traditional analytical approaches. Therefore, numerical models
have been developed to explore the properties of the solar 
wind-magnetosphere-ionosphere coupling. It has been demonstrated that this kind of 
study is a natural match for MHD numerical simulations\cite{hu2007ionospheric}. 

Over the years, 
different global MHD models have been developed to study the cause-and-effect of 
the space weather: 1) Lax-Wendroff model \cite{ogino1992global}, 
2) FV-TVD model \cite{tanaka1994finite}, 3) OpenGGCM model \cite{raeder1995structure}, 
4) GUMICS-3 model \cite{janhunen1996gumics}, 5) LFM model \cite{lyon2004lyon}, 
6) BATS-R-US model \cite{powell1999solution}, and 7) PPMLR-MHD model \cite{hu2007ionospheric}. 
The PPMLR-MHD model is a new global MHD 
model, which achieves high order spatial accuracy and low numerical dissipation 
compared to other existing models. By applying the piecewise parabolic method, the PPMLR-MHD algorithm has an accuracy of a third order in space, which enables the numerical model to present physical solutions even using relatively larger grid spacing. However, the parabolic interpolation is more complex than other interpolations, i.e. the linear interpolation, and therefore the computation amount will cost more time. Besides, this interpolation advancement also requires more communication data transfer. These problem together pose big challenge to develop a highly efficient fast implementation.

The development of the General-Purpose GPU technology in recent years has made 
big progress in the field of high performance computing. Due to the massive 
parallelism nature of GPUs, researchers are now able to solve large scale problems at a 
much faster speed while using less power. As of now 
NVIDIA has released its 4th-gen CUDA devices capable of running 
thousands of parallel threads simultaneously. However, 
GPU programming is very different from that of CPU as it has a very 
specific architecture, one needs to carefully design the architecture of the 
software to maximize the performance. Meanwhile, since GPU has 
its own dedicated memory, concerns also need to be addressed in minimizing the 
overhead of data transfer.

In this work, we present a GPU implementation of the PPMLR-MHD model. Due to the computation and data transfer intensive nature of this model, special attention has been paid to alleviate the overhead during simulation. We discuss 
the parallel problem partition, followed by the adaptation design 
scheme for taking advantage of the latest NVIDIA GPUs. We also demonstrate the techniques adopted to alleviate data transfer overhead using GPU direct techniques. We scale our implementation to hundreds of processes and solve the MHD simulation problem in a 
very efficient manner to meet the real time requirements for space weather 
forecast.

\section{The PPMLR-MHD Method}
\label{sec:method}

The global numerical model for Earth's space environment, especially for the 
magnetosphere  is based on the magnetohydrodynamic (MHD) description of plasma.
The conservational form of the 3-Dimensional MHD equations is listed as 
follows:

\begin{align*}
\frac{\partial\rho}{\partial t} &+ \nabla\cdot(\rho\bold{v}) = 0 \\
\raisetag{17pt}\frac{\partial(\rho\bold{v})}{\partial t} &+ \nabla\cdot(\rho\bold{v}\bold{v} +
p^{\ast}\bold{I} - \frac{1}{\mu_{0}}\bold{B}^{'}\bold{B}^{'}) =
(\nabla\times\bold{B}^{'})\times\bold{B_{d}}\\
\frac{\partial\bold{B}^{'}}{\partial t} &+ \nabla\cdot(\bold{v}\bold{B}^{'} -
\bold{B}^{'}\bold{v}) =
\nabla\times(\bold{v}\times\bold{B_{d}}) - \bold{v}\nabla\cdot\bold{B}^{'}\\
\begin{split}
\frac{\partial E}{\partial t} &+ \nabla\cdot[(E + p^{\ast})\bold{v} -
\frac{1}{\mu_{0}} (\bold{v}\cdot\bold{B}^{'})\bold{B}^{'}] = \\
&\;\quad\bold{v}\cdot[(\nabla\times\bold{B}^{'})\times\bold{B_{d}}]
+ \bold{B}^{'}\cdot[\nabla\times(\bold{v}\times\bold{B_{d}})]
\end{split}
\end{align*}
where
\begin{align*}
\bold{B}^{'} &= \bold{B} - \bold{B_{d}}, p^{\ast} = p + 
\frac{B^{'2}}{2\mu_{0}},\\
E &= \frac{p}{\gamma - 1} + \frac{1}{2}\rho v^{2} + \frac{B^{'2}}{2\mu_{0}},
\end{align*}
$\rho$ is the density, $p$ is the plasma
pressure, $\bold{v}$ is the flow velocity, $\bold{B}, \bold{B_{d}}, 
\bold{B^{'}}$ are the
magnetic field, the Earth's dipole field, and the difference
between the two fields respectively, $\mu_{0} = 4\pi\times 10^{-7}\text{H}\cdot\text{m}^{-1}$ 
is the permeability of vacuum and $\gamma = 5/3$ is the adiabatic index. In 
order to improve the accuracy in the calculation of the magnetic field, 
$\bold{B^{'}}$ is evaluated instead of $\bold{B}$ during the simulations, for 
the Earth's dipole field could be very large when closer to Earth. But it is 
noted that $\bold{B^{'}}$ does not have to be small compared with 
$\bold{B_{d}}$.

A so-called piecewise parabolic method with a
Lagrangian remap (PPMLR)-MHD algorithm, developed by Hu etc.~\cite{hu2007ionospheric}, 
is applied to solve these equations.
It is an extension of the Lagrangian version of the Piecewise
Parabolic Method (PPMLR) developed by Collela \& Woodward~\cite{colella1984piecewise} 
to MHD. In this PPMLR-MHD algorithm,
all variables ($\rho$, $\bold{v}$, $\bold{B^{'}}$ and $p$) are defined at the
zone centers as volume averages, and their spatial distributions can be obtained 
by a parabolic interpolation which is piecewise
continuous in each zone. When a characteristic method is used to calculate the 
local values of the variables at the
zone edges, they are updated first in the Lagrangian coordinates to the next 
time step, and then the
results are remapped onto the fixed Eulerian grids by solving the corresponding 
advection equations.

For the closure of the field-aligned current, an ionosphere shell, setting at r 
= 1.017$R_{E}$ (110 km altitude) is integrated in the simulation model under the 
electrostatic assumptions, which is connected with the magnetospheric inner 
boundary, setting at r = 3$R_{E}$ by dipole field lines between them. An 
electrostatic potential equation is solved in the ionosphere, and the solved 
potential is mapped to the magnetospheric inner boundary as a boundary condition 
for magnetospheric flows.

More numerical details of the simulation model can be found in 
\cite{hu2007ionospheric}, and a number of studies of the solar 
wind-magnetosphere-ionosphere interactions have been successfully carried out 
based on this model, and have been reviewed in 
\cite{wang2013magnetohydrodynamics}.

\section{The Hybrid Implementation}

\subsection{Task Partitioning}

The numerical domain of the PPMLR-MHD model is over a stretched Cartesian coordinate 
which takes the Earth as the origin and lets the x, y, and z axes 
point to the Sun, dusk, and the northward directions, respectively. The size of 
this domain extends from 30 $R_{E}$ to -100 $R_{E}$ along the Sun-Earth 
line and from -100 $R_{E}$ to 100 $R_{E}$ in y and z directions. It is 
discretized into 156 $\times$ 150 $\times$ 150 grid points: the grid is 
rectangular and nonuniform with the highest spatial resolution of 0.4 $R_{E}$ 
near Earth. To accommodate the large simulation volume and long simulation 
time, we parallelize the model to be scalable from several to hundreds of processes
, and partion the domain in all three directions. For the purpose of load balance, each 
subdomain usually contains the same number of grid points. 
Since the subdomain 
that contains the Earth must stay in a single numerical grid,
the number of processes in the y and z direction must be odd (1, 3, 5, etc.). 
In addition, one more process is required to solve the ionospheric potential equation. 
It is worthy to point out that the 3-Dimensional MHD equations on every grid 
in each subdomain is divided into three 1-Dimensional equations, which are independent
in the other two directions when solving these one dimensional equations.
As 156 is divisible by 3, 4 and 6, we employ the process configuration of 
3 $\times$ 1 $\times$ 1, 3 $\times$ 3 $\times$ 3, 4 $\times$ 3 $\times$ 3, 
6 $\times$ 3 $\times$ 3, 4 $\times$ 5 $\times$ 5, 6 $\times$ 5 $\times$ 5, 
plus one more process, the final number of processes are 4, 28, 
37, 55, 101, 151, respectively.

\subsection{Optimisation Analysis and Design for GPU}

GPU is different in hardware architecture compared to CPU in that it focuses 
more on the part of parallel computation in large scale. 
To maximize the computational power of the GPU, attentions have to be paid at the 
design phase of the GPU code implementation. This section first describes the 
general optimisation methods in GPU programming and then discusses the high 
performance design considerations adopted in our PPMLR-MHD implementation. 
In this work we employ NVIDIA's CUDA as our GPU programming tool, therefore the 
optimisation terminologies demonstrated are limited to CUDA. 

\subsubsection{Overview of Optimisation Methods}

In essence, GPU is a massively parallel device which is capable of running tens 
of thousands of tasks simultaneously. Similar to CPU, the basic GPU execution 
unit is termed as GPU thread. To achieve high efficiency on GPU, it is required 
that the GPU threads launched are well past the number of executable threads 
(tens of thousands) so that hanging and ready threads can be switched back and 
forth dynamically to hide load/store instruction latency. The general 
optimisation methods include (but not limited): a) increasing 
the occupancy (\cite{nvidia2008programming, cuda2013best}), which is defined as 
the ratio between the active executing threads and the maximal executable 
threads on GPU streaming multiprocessor (SMX), b) employing coalesced memory 
access pattern (adjacent threads access adjacent memory addresses in short)  for 
global memory read/write, c) shared memory tuning for reusable data access, d) 
atomic optimisation \cite{guo2015efficient} for superposition that involves 
small scale  memory conflict (no greater than 16 threads per memory address), e) 
warp shuffle optimisation for reduction, f) streaming for increasing 
multiprocess GPU resources utilization.

\subsubsection{High Efficiency Design}

\begin{figure}[htbp]
\centering
\vspace*{-0.07in}
\includegraphics[width=0.48\textwidth]{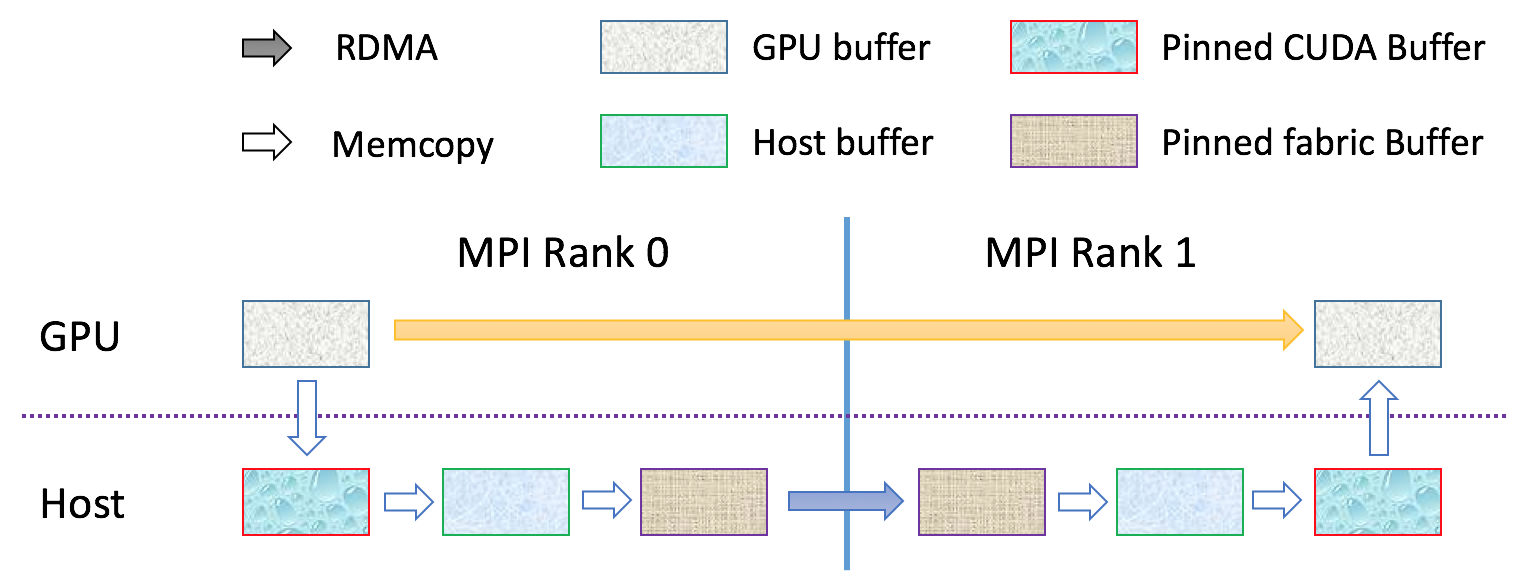}
\caption{Comparison between traditional MPI and CUDA-aware MPI. The CUDA-aware 
MPI directly transfers data via RDMA while avoiding unnecessary data movement 
compared to the traditional MPI.}
\label{fig:cuda-mpi}
\end{figure}

Our goal is to achieve high 
efficiency, high performance while maintaining high accuracy. In an effort to 
achieve this goal, we concentrate our effort on the following four aspects in 
designing the implementation: a) Adopt non-uniform spacing strategy in choosing 
numerical grids to reduce unnecessary computation. In our PPMLR-MHD model, a 
uniform mesh is laid out in the near-Earth domain within 10$R_{E}$, and the grid 
spacing outside increases according to a geometrical series of common ratio of 
1.05 along each axis. b) Reorganize the data so that it's GPU memory efficient. 
The memory-bound nature of the PPMLR-MHD method makes it very important to 
efficiently access the data as the whole simulation process consists of tens of 
thousands of time steps. In each step, the boundary input is obtained from 
neighbouring MPI processes and employed only once to solve the equations 
described in section \ref{sec:method}. As a result, GPU's shared memory is no 
longer needed for optimisation which makes global memory access with coalesced 
pattern very important. In addition, each solving process contains only basic 
arithmetic, no reduction or memory conflict is involved, therefore warp-shuffle 
and atomic optimisation are not helpful in our case. c) Employ streaming strategy to 
make the most of GPU resources. As kernels in different GPU streams can make use 
of the GPU parallelly, we make every grid execute in a specific stream. d) Make 
use of CUDA-aware MPI to efficiently transfer data between MPI ranks. In 
traditional MPI programming, the data computed on GPU has to be transferred back 
into CPU's host memory and then sent to responding MPI processes via MPI 
message. This is a big overhead as unnecessary data transfer is totally a waste of 
time. In our implementation, each grid needs to obtain boundary data from 
neighbouring processes in each step and there are more than thousands of steps 
performed, which makes the data transfer even more severe. To address this 
problem, we take advantage of the CUDA-aware MPI. Besides the capability of 
transferring data pointers pointed to host memory, the CUDA-aware MPI also takes 
care of GPU's device memory, which significantly reduces the unnecessary data 
movement by directly move data between GPUs by taking advantage of network 
adapter's RDMA capability. Figure \ref{fig:cuda-mpi} illustrates the data 
movement difference between traditional MPI and CUDA-aware MPI. The long solid 
arrow in the upper part of the figure shows the data flow of the CUDA-aware MPI, 
the rest of the arrows shows the data flow of the traditional MPI.

\section{Performance and Scalability Study}

This section discusses the performance and scalability of the GPU PPMLR-MHD 
implementation. To begin with, we study the scalability of the simulation, which 
gives a general performance picture of our MHD method. We then discuss more 
specific topics including single process performance and the transfer time of 
each process in different execution configurations. 

\subsection{Testing environment}

We test our implementation on  the Titan Supercomputer at the Oak Ridge National Laboraty (ORNL), the 
hardware and software configuration of each computing node is shown in Table 
\ref{tb:testconfig}, each node is equipped with two 10-core CPUs as well as one NVIDIA GPU.

\begin{table}[htbp]
\centering
\caption{Compute Node Configuration}
\label{tb:testconfig}
\setlength\extrarowheight{3pt}
\begin{tabular}{l | l}
\hline
CPU & AMD Opteron\texttrademark\ 6274 (Interlagos) \\
\hline
GPU & NVIDIA Tesla K20x \\
\hline
Host Max Memory Bandwidth & 51.2 GB/s \\
\hline
GPU Max Memory Bandwidth & 250 GB/s \\
\hline
Network & 40 Gb/s Infiniband  \\
\hline
Memory & 32 GB \\
\hline
OS & Cray Node Linux (CNL) \\
\hline
CUDA Version & 7.5 \\
\hline
\end{tabular}
\end{table}

In the case of the PPMLR-MHD simulation, the input scale depends on the 
resolution of the numerical grids, which is predetermined in the design phase 
of this method. In other words, the input scale is fixed. Therefore, we focus 
on the strong scalability of the PPMLR-MHD simulation. We test our 
implementation with  4, 28, 37, 55, 101, 151 processes, respectively, and then look 
into detailed performance results on both computation and data transfer.  The 
reason we choose these irregular numbers is that the PPMLR task 
partition scheme has its own constraint: the number of ranks in the y and z 
direction must be odd and the same as each other, plus one more MPI rank is used 
for calculating the boundary update information. To perform the simulation, we 
start with a somewhat arbitrarily prescribed initial state. In the domain of x 
\(\leqslant\) 15$R_{E}$, \textbf{\textit{B'}} (defined in section \ref{sec:method}) 
is produced by the image of Earth's dipole located at (x, y, z) = (30, 0, 0)$R_{E}$, 
and the initial distribution of plasma density and temperature is spherically symmetric.
Finally, on the right of x = 15$R_{E}$, a uniform solar wind and a 
uniform interplanetary magnetic field (IMF) are assumed. The simulation will 
continue until a steady-state magnetosphere is reached.

\subsection{Scalability and Accuracy study}

The accuracy and utility of space weather forecasts depend heavily on a 
thorough knowledge of the Sun-Earth system. The PPMLR-MHD model is designed to 
have high order spatial accuracy and low numerical dissipation. To further 
improve the accuracy in the calculation of the magnetic field, we have 
subtracted the Earth's dipole field from the total, and only the deviation 
field \textbf{\textit{B'}} is evaluated during the calculations. For the 
simulation, a given numerical accuracy can be achieved by using a corresponding 
numerical grids partition. In this work, we choose a relatively middle numerical 
grids (\(156 \times 150 \times 150\)) which is most frequently used in our daily 
working environment.

Since the overall problem for a given accuracy can be considered as fixed for a 
given accuracy, we discuss the strong scalability of the problem and omit the 
study of weak scalability. The strong scalability of the PPMLR-MHD method is 
determined by the size of the numerical grids. In our experiment configuration, 
the numerical grids is chosen to be \(156 \times 150 \times 150\). To 
facilitate implementation, we add extra constraint to our task partition 
strategy: the number of MPI ranks in the y and z axis must be the same and they 
are required to be odd numbers. PPMLR-MHD employs an iterative method to solve 
the MHD equations, in each iteration of the simulation, each grid (computed by a 
single MPI rank) needs to exchange its boundary data from all of its neighbours, 
the total amount of data required to be exchanged (TDE) is proportional to the 
partition choice applied. The amount of data exchanged in z direction can be 
represented as \(c \times n_x \times n_y \times (n_z - 1)\) where c represents 
a certain constants and \(n_x, n_y, n_z\) is the number of MPI ranks in the x, 
y, z dimensions, respectively, x and y direction can be calculated in the same 
way, there fore, TDE can be calculated as:

\begin{equation} 
\label{eq1}
\begin{split}
TDE & \propto n_x \times n_y \times (n_z - 1) \\
 & + n_x \times (n_y - 1) \times n_z \\
 & + (n_x - 1) \times n_y \times n_z
\end{split}
\end{equation}

\begin{figure}[htbp]
\centering
\vspace*{-0.07in}
\includegraphics[width=0.5\textwidth]{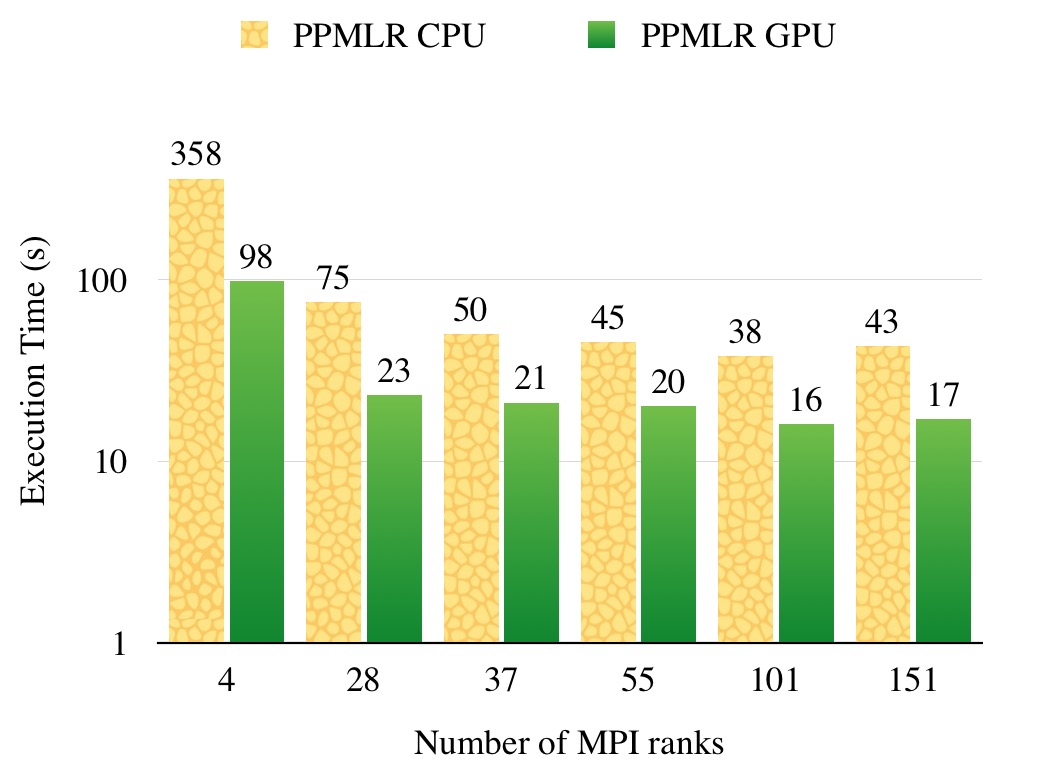}
\caption{Execution time of simulations for 100 iterations in different scale 
configuration.}
\label{fig:Hundred}
\end{figure}

From equation \ref{eq1}, the total number of MPI ranks has positive relation to 
the total amount of exchanged data, involving more GPUs also brings more data 
transfer overhead. Besides, GPU requires problem to be big enough in order to 
take advantage of its massive parallel power. A compromise has to be made 
between the computation resources and the data exchange overhead. To 
demonstrate, we simulate our implementation for 100 iterations since 
the total amount of execution time required for several execution configuration 
is too long. Figure \ref{fig:Hundred} shows the simulation results.

As illustrated in the figure, in every chosen execution configuration, our GPU 
PPMLR-MHD implementation outperforms the CPU counterpart (2.5 to 3.5 times 
faster). The best performance comes from the execution configuration of 101 MPI 
ranks, this result meets our expectation, because more processes bring not only more 
computational resources but also more data exchange overhead, the execution 
configuration of 101 processes is the balance point between these two performance 
related factors.

\subsection{GPU performance study}

The computing capability of GPU plays a key role for us to achieve real time 
simulation. This section studies the performance contribution of GPU. To 
begin with, we compare our GPU implementation with a CPU counterpart 
which is highly optimised for spatial-temporal locality memory access. To 
accurately measure the computation performance, we randomly choose several single steps 
and record the computation time from each MPI rank, we then average the sum of 
all records and use the final result to perform comparison. Since the 
computing workload for each step is predetermined and thus fixed, each
step is supposed to take approximately the same amount of time. Our results
give a direct comparions of the code performance on GPU and CPU.

Although parallelization is well considered in the design process of the PPMLR-MHD 
implementation, the memory-intensive nature of the algorithm makes it more memory bandwidth 
limited when considering the expected performance. In theory, the PPMLR-MHD 
GPU implementation's maximum achievable speedup (MAS) over the CPU counterpart can 
be expressed as: 

\[ MAS = \frac{GPU\ Memory\ Bandwidth}{Host\ Memory\ Bandwidth} \]

\begin{figure}[htbp]
\centering
\vspace*{-0.07in}
\includegraphics[width=0.5\textwidth]{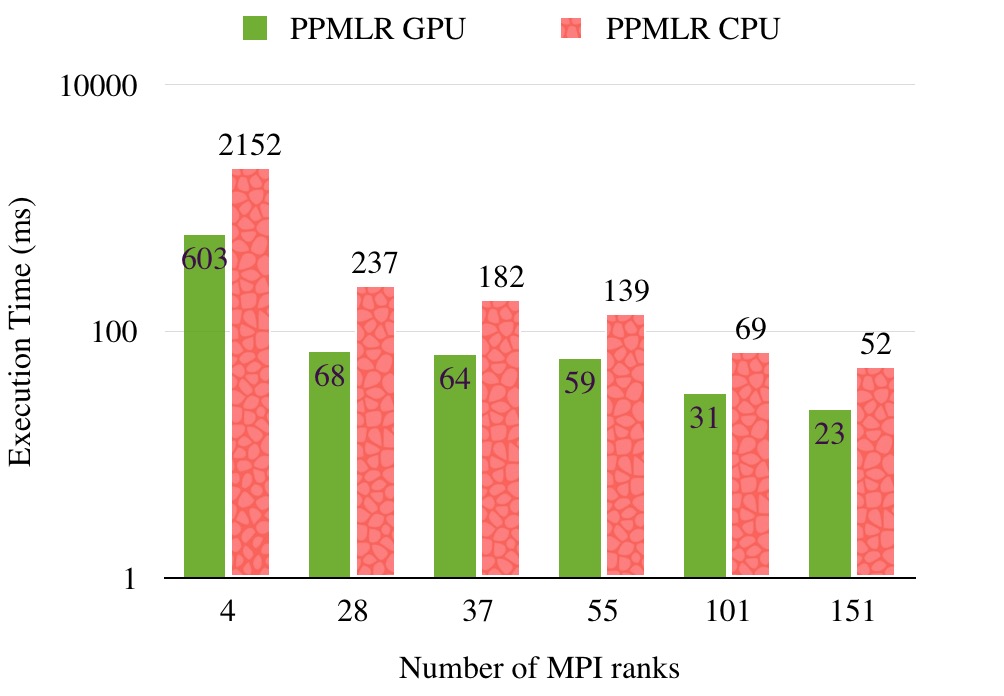}
\caption{Average execution time of each step in different scale configurations. 
The number in the x axis stands for the number of processes used in launching MPI 
executables.}
\label{fig:GPU-CPU}
\end{figure}

From the hardware configuration listed in Table \ref{tb:testconfig}, the maximum 
expected speedup should be MAS = 250 GB/s / 52 GB/s \(\approx\) 4.88 
in our testing environment.  
Figure \ref{fig:GPU-CPU} illustrates the execution time of a 
single step in different scale configurations.

As can be seen in the figure, the employment of GPU has significantly improves 
the performance of the PPMLR simulation. The GPU version has achieved up to 
3.57x speedups (compared to the CPU counterpart) in the configuration of 4 MPI 
ranks. This result is very close to the MAS theoretical peak of 4.88 (73.2 \% of 
the GPU bandwidth has been employed). As the number of MPI ranks increases, the 
effect of GPU acceleration is less significant, this is due to the fact that 
the increasing of parallel processes results in the decreasing of the amount of workload in 
each MPI rank. When the workload is below a certain amount, GPU is underutilize 
thus unable to achieve peak performance. 

\subsection{Data transfer time study}

Heterogeneous computing devices such as GPU have their own dedicated memory. Data 
that are used as input to execute are required to reside on 
GPU's device memory, as the host memory is not accessible from GPU streaming 
multiprocessors. For many traditional MPI-based scientific applications such as PPMLR-MHD, 
this causes a major problem in using multiple GPUs since the computed results must be 
transferred back and forth from GPU memory to 
CPU's host memory in order to update boundary data on the numerical grids.
Arrows in the lower part of Figure \ref{fig:cuda-mpi} demonstrates the data flow from
one GPU to another using traditional MPI. As can be seen, except for 
the necessary RDMA operation, there are 6 more memory copy operations involved 
in a single data transfer between two MPI ranks. Obviously, the redundant 6 
memory copy operations per MPI rank pair causes a bandwidth overhead that is 
non-ignorable in achieving high performance.

To reduce the data transfer overhead, we employ the CUDA-aware MPI. Through this 
technology, the GPU's device memory can be directly sent to/received from the 
MPI api, combined with Remote Direct Memory Access (RDMA) technology, buffers 
can be directly sent from GPU memory to a network adapter without staging 
through host memory, as shown in the upper part of Figure~\ref{fig:cuda-mpi}. 
Therefore, compared to a CPU counterpart using traditional 
MPI implementation, our PPMLR-MHD GPU version is supposed to take approximately 
the same amount of time in transferring data. We predict that

\[ CPU\ data\ transfer\ time \approx GPU\ data\ transfer\ time \]

To check our prediction, we randomly choose a single simulation step from both 
PPMLR-MHD GPU implementation and the CPU counterpart, we record the total amount 
of time taken for sending as well as receiving data for all MPI ranks involved, 
we then average them and use the result as the comparison input. Since the 
amount of data being transferred is exactly the same for every single step, the 
choice of the experiment design is supposed to demonstrate the accurate data 
transfer overhead difference between the CPU and GPU implementation.

\begin{figure}[htbp]
\centering
\vspace*{-0.07in}
\includegraphics[width=0.5\textwidth]{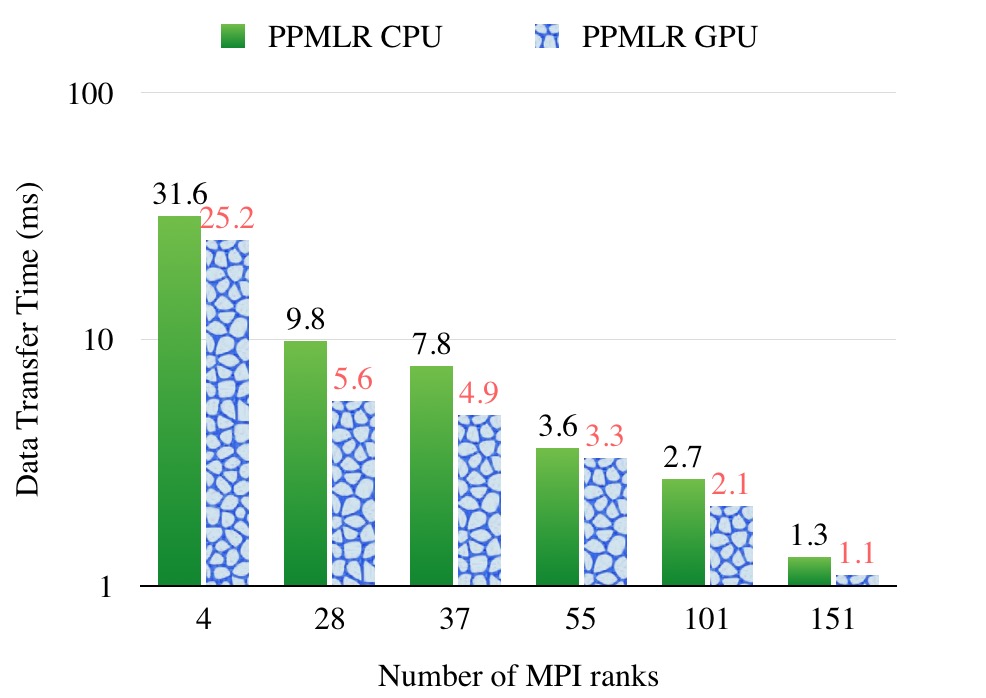}
\caption{Average data transfer time of each step in different scale 
configurations. The number in the x axis represents the number of processes used in 
launching MPI executables.}
\label{fig:data-transfer}
\end{figure}

Figure \ref{fig:data-transfer} shows the comparison of the data transfer time of
the PPMLR GPU implementation versus that of the CPU counterpart in different execution 
configurations. As presented in the figure, the PPMLR GPU implementation takes 
slightly less time in transferring data compared to the CPU counterpart. We 
attribute this result to the employment of the CUDA-aware MPI as well as GPU's 
high speed memory bus. CUDA-aware MPI helps our implementation to take advantage 
of the RDMA in transferring data among MPI ranks, as the CPU version also uses 
this technology, we expect the two implementations spend approximately the same 
amount of time in transferring data. The reason PPMLR GPU takes less time is due 
to the fact that GPU has higher speed bus, thus data from GPU's device 
memory reaches network adaptor quicker than that from host memory.

\section{Conclusion}

In this work, we present a GPU accelerated implementation of the PPMLR-MHD model
for space weather forecast. We significantly improve the code performance by taking 
advantage of the GPU technology as well as CUDA-aware MPI. By 
making careful choices in the design phase of the implementation, we are able 
to make the most of GPU's powerful compute capability and achieve near peak 
performance. The final implementation is scaled to up to 151 processes and the real 
time performance requirement is met.


\section*{Acknowledgment}
This research is supported in part by National Natural Science Foundation of 
China (Nos. 61440057,61272087 ,61363019 and 61073008), Beijing
Natural Science Foundation (Nos. 4082016 and 4122039), the Sci-Tech 
Interdisciplinary Innovation and Cooperation Team Program of the Chinese Academy 
of Sciences, the Specialized Research Fund for State Key Laboratories.

This research used resources of the Oak Ridge Leadership Computing Facility at 
the Oak Ridge National Laboratory, which is supported by the Office of Science 
of the U.S. Department of Energy under Contract No. DE-AC05-00OR22725.


%



\bibliographystyle{IEEEtran}
\bibliography{ref}

\end{document}